\documentclass[12pt]{iopart}
\usepackage{graphicx}

\newcommand{\req}[1]{(\ref{#1})}
\newcommand{\be}{\begin{equation}}
\newcommand{\ee}{\end{equation}}
\newcommand{\bea}{\begin{eqnarray}}
\newcommand{\eea}{\end{eqnarray}}

\newcommand{\bxi}{\mbox{\boldmath ${\xi}$}}

\newcommand{\bsigma}{\mbox{\boldmath ${\sigma}$}}

\newcommand{\odis}{\big<\hspace{-1mm}\big<}
\newcommand{\cdis}{\big>\hspace{-1mm}\big>}

\begin{document}
\title{Optimal static and dynamic recycling of defective binary devices}
\author{Damien Challet\dag\ and Isaac P\'erez Castillo\ddag} 
\address{\dag\ Nomura Centre for Quantitative Finance, Mathematical Institute, Oxford University, 24--29 St Giles', Oxford OX1 3LB, United Kingdom}
\address{\ddag\ Instituut voor Theoretische Fysica, Katholieke Universiteit Leuven,  B-3001 Leuven, Belgium}
\begin{abstract}
The binary Defect Combination Problem consists in finding a fully working subset from a given ensemble of imperfect binary components. We determine the typical properties of the model using methods of statistical mechanics, in particular, the region in the parameter space where there is almost surely at least one fully-working subset. Dynamic recycling of a flux of imperfect binary components leads to zero wastage.
\end{abstract}
\ead{\tt challet@maths.ox.ac.uk, Isaac.Perez@fys.kuleuven.ac.be}
\section{Introduction}
Producing fault-free devices such as computer processors so costly that only a few large companies can afford building and running new facilities. But even devices known to be fully working initially may fail a posteriori. Fault-tolerant computing tries to minimize the consequences of component failure by designing computer systems that continue to operate satisfactorily even in the presence of faults \cite{BWJohnson}. The majority of fault-tolerant designs involves partitioning a computer system into modules that act as fault-containment regions. Redundancy of these modules is then considered, so if one fails others can assume its function, optimizing reliability availability or efficiency.

While redundancy is expensive, components known to be imperfect are classified as useless and become cheap if not free, even though they can still be of some use. For instance, some devices with minor defects are still profitable, as faulty memory chips in answering machines \cite{IEEE}. Another example is the massive parallel computer Teramac \cite{teramac}, designed with devices with unknown status but connected with adaptive wiring so as to avoid the defects. A third strategy was presented in \cite{Ch02} by one of us, where it was noticed that devices are most often only partly defective and therefore one may combine them in such a way that their imperfections cancel. This is the essence of the Defect Combination Problem (DCP), which applies to both analog and binary components. While the analog problem was already addressed mathematically in \cite{Ch02}, the aim of the present paper is to solve its binary counterpart.

The paper is organized as follows: in the following section, section \ref{sec:problem_definition}, we present the problem and discuss briefly how to treat it by tools and concepts of statistical mechanics of disordered systems. The canonical ensemble approach is used in section \ref{sec:canonical} to analyze the typical properties of the model.  In section \ref{sec:simulations} we compare the analytical work with numerical simulations. Section \ref{sec:recycling} is devoted to the flux recycling problem.
\section{Model definitions}
\label{sec:problem_definition}
We assume that each device is able to perform $P$ different functions, numbered by $\mu=1,\cdots,P$. The manufacturing process is such that each function is either permanently defective with probability $\phi$ or working with probability $1-\phi$.\footnote{This is not unrealistic: the quality of an electronic chip is determined by the local level of impurities of the silicon wafer from which it is made. Local fluctuations of the impurities density can cause a function to be defective.}. The DCP consists in extracting from an ensemble of $N$ devices a subset such that the defects compensate optimally.
 
More precise, let us denote with Ising variables $\xi_i^\mu\in\{-1,+1\}$ whether the function $\mu$ of the component $i$ is defective ($\xi^\mu_i=-1$) or not ($\xi^\mu_i=1$). This means that the manufacturing process is summarized by 
\begin{equation}
P(\xi^\mu_i)=\phi\delta(\xi^\mu_i+1)+(1-\phi)\delta(\xi^\mu_i-1)\,,
\label{probxisphi}
\end{equation}
which assumes that the state of the functions is determined independently at the time at which the device is being made.\footnote{This is akin to assume that the local fluctuations of impurity density in the wafer example are uncorrelated.}
To identify whether a component $i$ belongs to a specific subset we introduce the boolean variables $\sigma_{i}\in\{0,1\}$ such that if $\sigma_i=1$ the component belongs to a given subset of zero otherwise. Every possible subset out of the ensemble is fully determined by a vector $\bsigma=(\sigma_1,\ldots,\sigma_N)$.  The binary DCP is defined as the search for a combination such that the majority of its components gives the correct answer for all the functions, in which case
\begin{equation}
\label{condition}
\sum_{i=1}^N\xi^\mu_i\sigma_i\geq0,\quad \forall \mu=1,\ldots,P\,.
\end{equation}
Conditions (\ref{condition}) are also called majority vote in the fault-tolerant computer literature. A simple inspection of the set of inequalities (\ref{condition}) indicates that a phase transition is expected. Indeed, it is clear that when $N\gg P$ there is a large number of subsets out of the possible $2^N$ satisfying the above conditions. When $P$ increases the number of these subsets decreases and finding configurations that satisfy the majority vote (\ref{condition}) is increasingly difficult; at some point finding perfect subsets is not possible any more. Therefore, one expects the existence two phases: a fault-free one where perfect subsets exist ($\alpha=P/N<\alpha_c$) and a imperfect one where conditions \req{condition} $\alpha>\alpha_c$ where the best one can do is minimizing the number of unsatistifed conditions, that is, the number of faulty functions, denoted by $C$.

We define $C(\bsigma)$, the number of unsatisfied conditions in subset $\bsigma$ as
\begin{equation}
{\cal C}(\bsigma)=\sum_{\mu=1}^P\Theta \left(\kappa-\frac{1}{\sqrt{N}}\sum_{i=1}^N\xi^\mu_i\sigma_i\right)\,,
\label{cost}
\end{equation}
where $\kappa$ is a confidence threshold ($\kappa=0$ corresponds to the majority vote). If one rewrites the default distribution function as 
\begin{equation}
P(\xi^\mu_i)=\frac{1+\gamma}{2}\delta(\xi^\mu_i+1)+\frac{1-\gamma}{2}\delta(\xi^\mu_i-1),\quad \gamma=m/\sqrt{N}
\label{ probxis}
\end{equation}
the similarity between the binary DCP and the optimal capacity problem of Ising neural networks \cite{Ga88,Ga88b} is evident. More precisely, the binary DCP is equivalent to the optimal capacity problem with $J=0,1$ synaptic couplings and biased patterns introduced in \cite{Gu90}\footnote{It seems however that there is an inconsistency in \cite{Gu90} in the way  the order parameters scale with the system size that has been unnoticed. Following their notation, the presence of bias in the patterns implies that one must introduce the order parameter $(1/\sqrt{N})\sum_{i=1}^NJ_{i}^\alpha$ while the diagonal part of the spin glass overlap gives $(1/N)\sum_{i=1}^NJ_{i}^\alpha$, thus having essentially the same parameter but scaling differently with the system size. It may be even possible that this was the source of discrepancy between different models in the sparse coding limit \cite{Gu90,Nadal1991a,Brunel1994}. A forthcoming work will address this question in details \cite{CPremark04}. We note here that this inconsistency is cured as soon as one rescales the bias parameter with the system size as in \req{ probxis}.}.

Whereas any combination can be considered in the above problem, the constrained DCP restricts the choice of combinations to those comprising a fixed number $\sum_{i=1}^N\sigma_i$ of components. The technological justification for this is that an actual implementation of the DCP would be made easier by building in advance boards designed for receiving a fixed number of components.
\section{Canonical approach}
\label{sec:canonical}
We shall proceed similarly as in \cite{Ga88,Ga88b,Kr89}. In the canonical ensemble the typical properties of the unconstrained DCP are fully described by the partition function
\begin{equation}
{\cal Z}(\beta)=\sum_{\bsigma}e^{-\beta{\cal C}(\bsigma)}\,,
\label{partition}
\end{equation}
with $\beta=1/T$ the inverse temperature. The free energy $f$ then reads
\begin{equation}
 f(\beta)=-\lim_{N\to\infty}\frac{1}{\alpha N \beta}\odis\log {\cal Z}(\beta)\cdis_{\bxi}\,,
\label{ free_energy}
\end{equation}
where $\odis [\cdots]\cdis_{\bxi}$ denotes the average with respect to $P(\bxi)$. We are interested in the zero temperature limit where the free energy $f$ corresponds to the fraction of erroneously implemented functions whilst the entropy, defined as
\begin{equation}
s(\beta)=\alpha\beta^2\frac{\partial  f(\beta)}{\partial \beta}\,,
\label{ entropy}
\end{equation}
is the logarithm of the number of solutions to the DCP. Nevertheless it is interesting to point out the behavior of this model with the temperature, similar to the Random Energy Model \cite{Kr89,De81}. In the faulty-free regime ($\alpha<\alpha_c$) the entropy is positive at any temperature while the free energy vanishes at zero temperature since there are perfect subsets. In this regime the Replica Symmetric (RS) approximation is indeed a very good approximation (if not exact) at any temperature. The critical point $\alpha_c$ coincides with the cancellation of the entropy at zero temperature, because there are no more perfect subsets. However for $\alpha>\alpha_c$  there exists a critical temperature $T_c$, called freezing temperature, below which RS is broken (the entropy becomes negative). A One-Step Replica Symmetry Breaking (RSB) calculation reveals that for $T\leq T_c$ the RS entropy becomes zero while the RS internal energy (fraction of errors) freezes to its value at $T_c$ for $T\leq T_c$\footnote{This discussion is based on a calculation in the canonical ensemble. One can also use the microcanomical ensemble as in \cite{Fontanari1993} in which a RS calculation is equivalent to the One-Step RSB calculation in the canonical one.}.

Starting from the expression (\ref{ free_energy}) and following standard procedures \cite{MPV} we write the free energy as
\begin{equation}
 f(\beta)=-\lim_{N\to\infty}\lim_{n\to0}\frac{1}{\alpha Nn \beta}\log\odis {\cal Z}^n(\beta)\cdis_{\bxi}\,,
\end{equation}
where we have used the replica approach based on the equivalence $\odis\log{\cal Z}\cdis=\lim_{n\to0}(\odis{\cal Z}^n\cdis-1)/n$, consisting in substituting the logarithm appearing in the equation \req{ free_energy} by an object much easier to average over the disorder. The $n$-th power in the partition function indicates that the same system has been replicated $n$ times, thus the name of replica. After some straightforward manipulations, the replicated and averaged partition function becomes
\begin{equation}
\odis{\cal Z}^n(\beta)\cdis_{\bxi}=\int\left[\prod_{\alpha,\beta=1}^n \frac{dq_{\alpha\beta}\widehat{q}_{\alpha\beta}}{4\pi i/N}\right]e^{N(G_1+G_2+G_3)}\,,
\label{ replicated_average_numberofsol}
\end{equation}
where we have defined the following macroscopic order parameters
\begin{equation}
q_{\alpha\beta}=\frac{1}{N}\sum_{i=1}^N\sigma_i^\alpha\sigma_i^\beta\,,\quad q_{\alpha\alpha}\equiv M_{\alpha}=\frac{1}{N}\sum_{i=1}^N\sigma_i^\alpha
\label{ order_parameters}
\end{equation}
and with the functions  $G_1$, $G_2$ and $G_3$ given by 
\begin{eqnarray}
G_1&=&\frac{1}{2}\sum_{\alpha,\beta=1}^nq_{\alpha\beta}\widehat{q}_{\alpha\beta}\label{ G1}\\
G_2&=&\alpha\log\int\left[\prod_{\alpha=1}^n\frac{dh_\alpha d\widehat{h}_\alpha}{2\pi }\right]\exp\left(i\sum_{\alpha=1}^n\widehat{h}_\alpha h_\alpha+im\sum_{\alpha=1}^n\widehat{h}_\alpha M_\alpha\nonumber\right.\\
&&\left.-\frac{1}{2}\sum_{\alpha,\beta=1}^n\widehat{h}_\alpha \widehat{h}_\beta q_{\alpha\beta}\right)\prod_{\alpha=1}^n\left[e^{-\beta}+(1-e^{-\beta})\Theta\left(h_\alpha-\kappa\right)\right]\label{ G2}\\
G_3&=&\log\sum_{\{\sigma^\alpha\}}\exp\left(-\frac{1}{2}\sum_{\alpha,\beta=1}^n\widehat{q}_{\alpha\beta}\sigma^\alpha\sigma^\beta\right)\label{ G3}\,.
\end{eqnarray}
In the thermodynamic limit ($N$, $P\to\infty$ at fixed $\alpha=P/N$) the expression (\ref{ replicated_average_numberofsol}) is evaluated by the steepest descent method and the free energy $f$ simply reads
\begin{equation}
f(\beta)=-\lim_{n\to0}\frac{1}{\alpha\beta  n}{\rm extr}_{\{q_{\alpha\beta},\widehat{q}_{\alpha\beta}\}}(G_1+G_2+G_3)\,.
\end{equation}
Within the RS ansatz the overlap parameters with two replica indexes are assumed to be invariant under the interchange of all replica indexes. We then write
\begin{eqnarray}
q_{\alpha\beta}&=&M\delta_{\alpha\beta}+q(1-\delta_{\alpha\beta})\\
\widehat{q}_{\alpha\beta}&=&\widehat{M}\delta_{\alpha\beta}+\widehat{q}(1-\delta_{\alpha\beta})\,.
\label{ order_parameters_RS}
\end{eqnarray}
Evaluation of the free energy $ f(\beta)$ and the entropy $s(\beta)$ in the RS ansatz gives
\begin{eqnarray}
  f_{{\rm RS}}(\beta)&=&-\frac{1}{2\beta\alpha}\big(M\widehat{M}+q\widehat{q}\big)-\frac{1}{\beta\alpha}\int Dt\log\left(1+e^{-\frac{\widehat{M}+\widehat{q}}{2} -t\sqrt{\widehat{q}}}\right)\nonumber\\
&&-\frac{1}{\beta}\int Dt\log\left[e^{-\beta}+(1-e^{-\beta}) H\left(\frac{\kappa+mM+t\sqrt{q}}{\sqrt{M-q}}\right)\right]
\label{ free_energy_RS}
\end{eqnarray}
and 
\begin{eqnarray}
\hspace{-1.5cm}s_{{\rm RS}}(\beta)&=&-\beta \alpha f_{{\rm RS}}(\beta)+\beta\alpha e^{-\beta}\int Dt\frac{1-H\left(\frac{\kappa+mM+t\sqrt{q}}{\sqrt{M-q}}\right)}{e^{-\beta}+(1-e^{-\beta}) H\left(\frac{\kappa+mM+t\sqrt{q}}{\sqrt{M-q}}\right)}\,,
\label{ entropy_RS}
\end{eqnarray}
with $Dt\equiv dt\,e^{-t^2/2}/\sqrt{2\pi}$  and $H(x)\equiv{\rm erfc}(x/\sqrt{2})/2$,  ${\rm erfc}(x)$ being the complementary of the error function.  The previous free energy (\ref{ free_energy_RS}) must be stationary with respect to $M,\widehat{M},q,\widehat{q}$. For the constraint model, stationary with respect to $M$ must not be imposed, and $M$ becomes a parameter that controls the relative size of the subset. The saddle-point equations read
\begin{eqnarray}
M&=&\int Dt\left(1+e^{\frac{\widehat{M}+\widehat{q}}{2} +t\sqrt{\widehat{q}}}\right)^{-1}\label{ fe1}\\
q&=&\int Dt\frac{1+\frac{t}{\sqrt{\widehat{q}}}}{1+e^{\frac{\widehat{M}+\widehat{q}}{2} +t\sqrt{\widehat{q}}}}\label{ fe2}\\
\widehat{M}&=&-2\alpha\int Dt\frac{(1-e^{-\beta})H_{,M}\left(\frac{\kappa+mM+t\sqrt{q}}{\sqrt{M-q}}\right)}{e^{-\beta}+(1-e^{-\beta}) H\left(\frac{\kappa+mM+t\sqrt{q}}{\sqrt{M-q}}\right)}\label{ fe3}\\
\widehat{q}&=&-2\alpha\int Dt\frac{(1-e^{-\beta})H_{,q}\left(\frac{\kappa+mM+t\sqrt{q}}{\sqrt{M-q}}\right)}{e^{-\beta}+(1-e^{-\beta}) H\left(\frac{\kappa+mM+t\sqrt{q}}{\sqrt{M-q}}\right)}\,,\label{ fe4}
\end{eqnarray}
with
\begin{eqnarray}
H_{,M}\left(\frac{\kappa+mM+t\sqrt{q}}{\sqrt{M-q}}\right)&=&\frac{1}{2}\frac{[\kappa+t\sqrt{q}-m(M-2q)]}{\sqrt{2\pi(M-q)^3}}e^{-\frac{(\kappa+mM+t\sqrt{q})^2}{2(M-q)}}\label{ H,m}\\
H_{,q}\left(\frac{\kappa+mM+t\sqrt{q}}{\sqrt{M-q}}\right)&=&-\frac{1}{2}\frac{[tM+\sqrt{q}(\kappa+mM)]}{\sqrt{2\pi q(M-q)^3}}e^{-\frac{(\kappa+mM+t\sqrt{q})^2}{2(M-q)}}\label{ H,q}\,\,.
\end{eqnarray}
We remark that equation (\ref{ fe3}) is not present in the constraint case. We will now study the different regimes.
\subsection{Faulty-free regime, $\alpha<\alpha_c$}
\begin{figure}[h]
\begin{center}
\includegraphics[width=0.5\textwidth,height=0.4\textwidth]{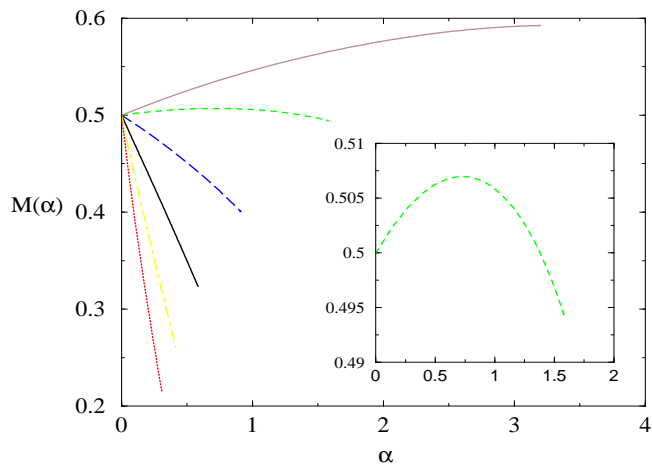}
\caption{Unconstraint case. Typical relative size $M$  versus $\alpha$ for $\kappa=0$ and $m=1.0,0.5,0.0,-0.5, -1.0$ and $-1.5$ (bottom to top). In the inset the case $m=-1.0$.}
\label{fig:M_unconstraint_anyalpha}
\end{center}
\end{figure}

At zero temperature the fraction of errors $f(\infty)$ becomes zero in this regime, while the entropy $s(\infty)$ is different from zero indicating that there exists perfect subsets. We have solved the saddle-point equations (\ref{ fe1})-(\ref{ fe4}) numerically at zero temperature (analytically this limit is calculated trivially) and for different values of the parameters $\alpha,m,\kappa$.
\begin{figure}[h]
\begin{center}
\includegraphics[width=0.7\textwidth,height=0.40\textwidth]{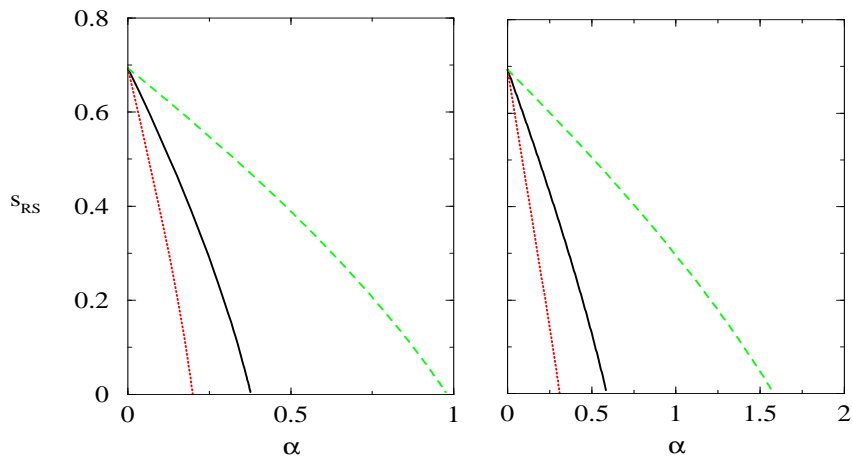}
\caption{Unconstraint case. Entropy $s_{{\rm RS}}(m,\kappa,\alpha)$ versus $\alpha$. Left panel: $\kappa=0$ and $m=1$ $\kappa=0$ a$-1.0$ (left to right). Right panel $\kappa=0.2$ and $m=1.0,0.0$ and $-1.0$ (left to right).}
\label{fig:unconstraint_anyalpha}
\end{center}
\end{figure}
At $\alpha=0$ the entropy is simply $s=\log (2)$, that is, there are $2^N$ perfect combinations. When $\alpha$ increases, the relative number of inequalities (\ref{condition}) increases and the number of perfect subsets decreases accordingly, diminishing the entropy as well. Note that as long as the entropy is finite, there is still an exponential number of perfect combinations. This behavior appears in figure \ref{fig:unconstraint_anyalpha} where we have plotted the entropy against $\alpha$ for different values of $m$ and $\kappa$ for the unconstraint case (the constraint case presents a similar behavior). We have also plotted how the typical size $M$ varies with $\alpha$. It presents typically a monotonic behavior depending on the value of $m$, but there is an interval where $M$ is non-monotonic (see inset in figure \ref{fig:M_unconstraint_anyalpha}).
A na\"\i ve explanation of this monotonic behavior would be as follows: let us first assume that with the same probability we may find defects or not ($m=0$). At $\alpha=0$, there are no constraints, hence all combinations have the same probability to be perfect and therefore $M=1/2$. As $\alpha$ increases it becomes more difficult to find perfect subsets and larger subsets are less likely to satisfy the majority vote. Consequently, the average size $M$ is reduced as $\alpha$ increases. Now for fixed $\alpha$ and as $m$ increases there are more defects and the large subsets are even less likely to satisfy the set of constraints (\ref{condition}) and consequently the average size $M$ becomes more reduced. For negative $m$ the opposite effect is observed for obvious reasons.
\subsection{Critical regime, $\alpha=\alpha_c$}
\begin{figure}[h]
\begin{center}
\includegraphics[width=0.4\textwidth,height=0.4\textwidth]{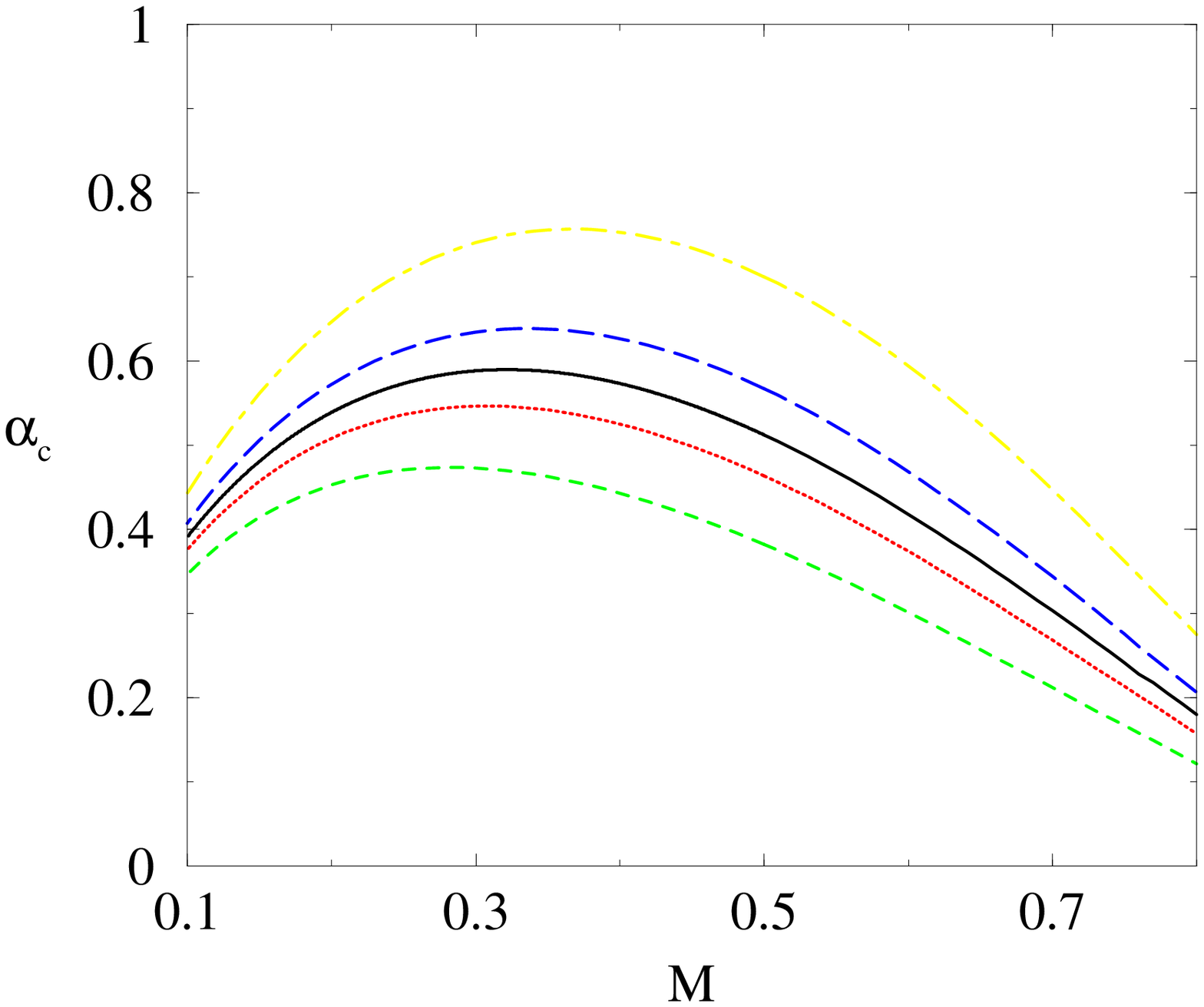}\quad 
\includegraphics[width=0.4\textwidth,height=0.4\textwidth]{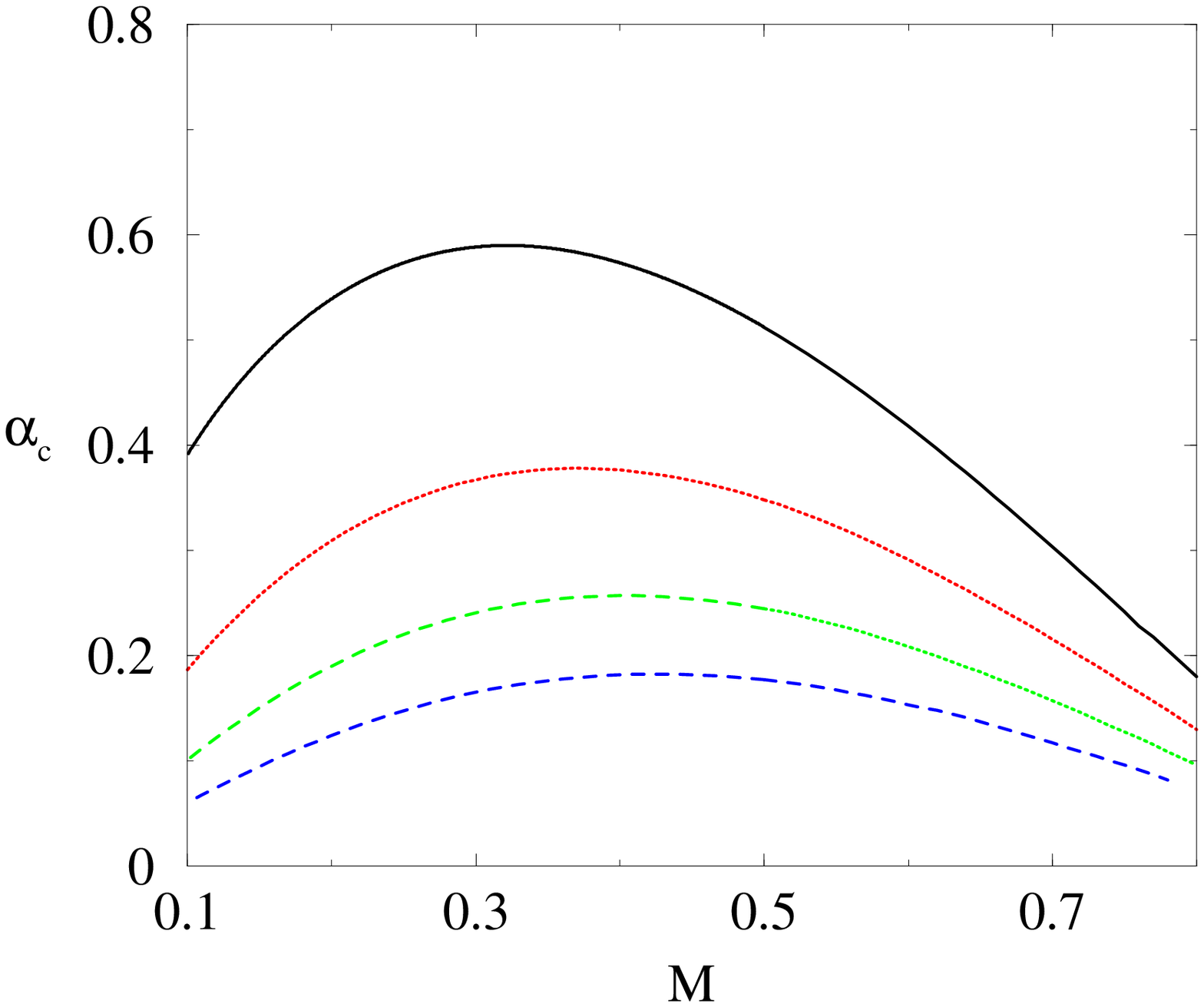}  
\caption{Constraint case. Left Panel: $\alpha_c$ versus relative size $M$ for $\kappa=0$ and for $m=-0.3,-0.1,0.0,0.1$ and $0.3$ (top to bottom). Right Panel: $\alpha_c$ versus relative size $M$ for $m=0$ and for $\kappa=0.0,0.2,0.4$ and $0.6$ (top to bottom)}
\label{fig:constraintZE}
\end{center}
\end{figure}

From figure \ref{fig:unconstraint_anyalpha} we see that the critical point is reached when the entropy becomes zero, {\it i.e.} there exists no more perfect subsets but one. 
\begin{figure}[h]
\begin{center}
\includegraphics[width=0.4\textwidth,height=0.4\textwidth]{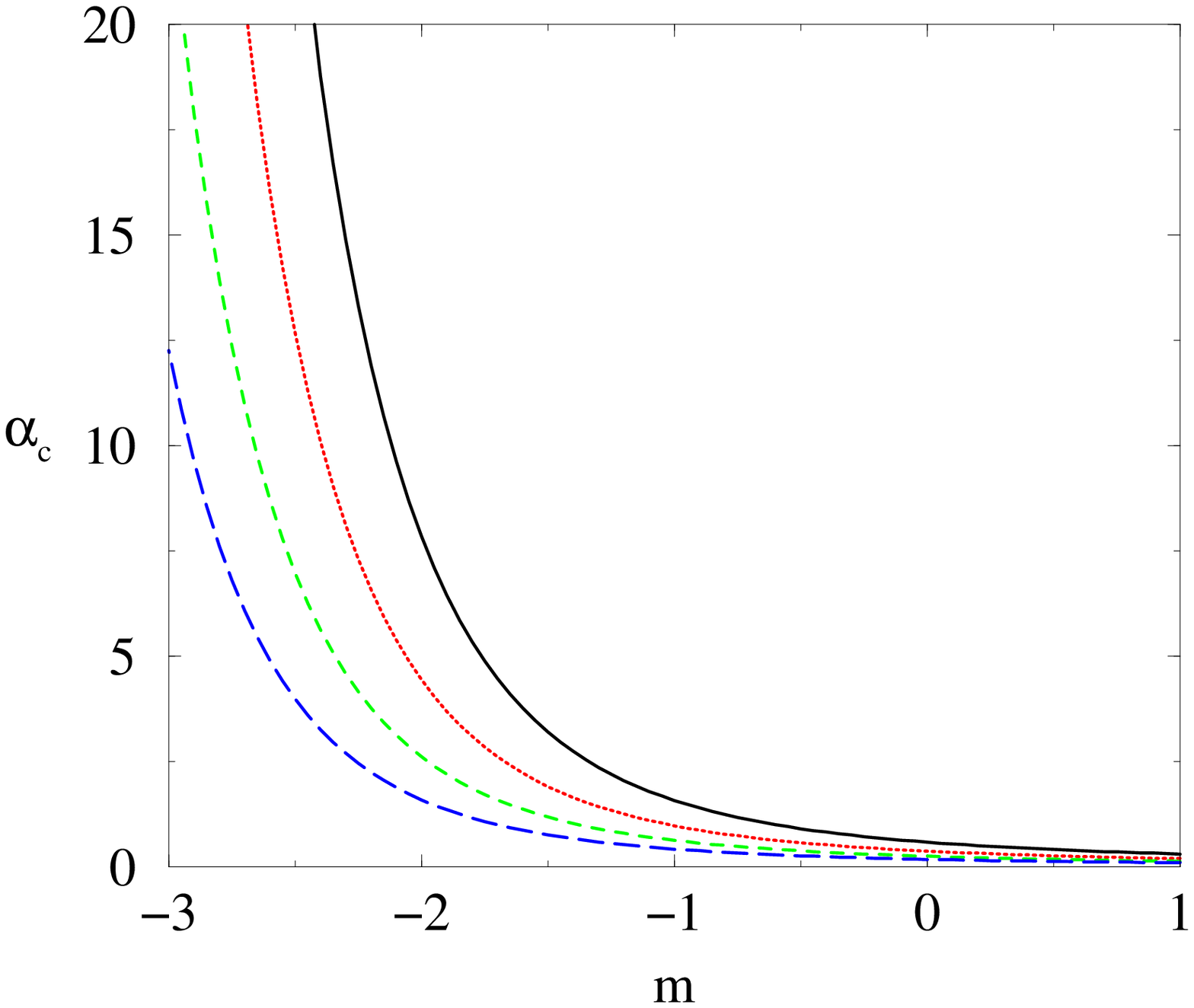}\quad
\includegraphics[width=0.4\textwidth,height=0.4\textwidth]{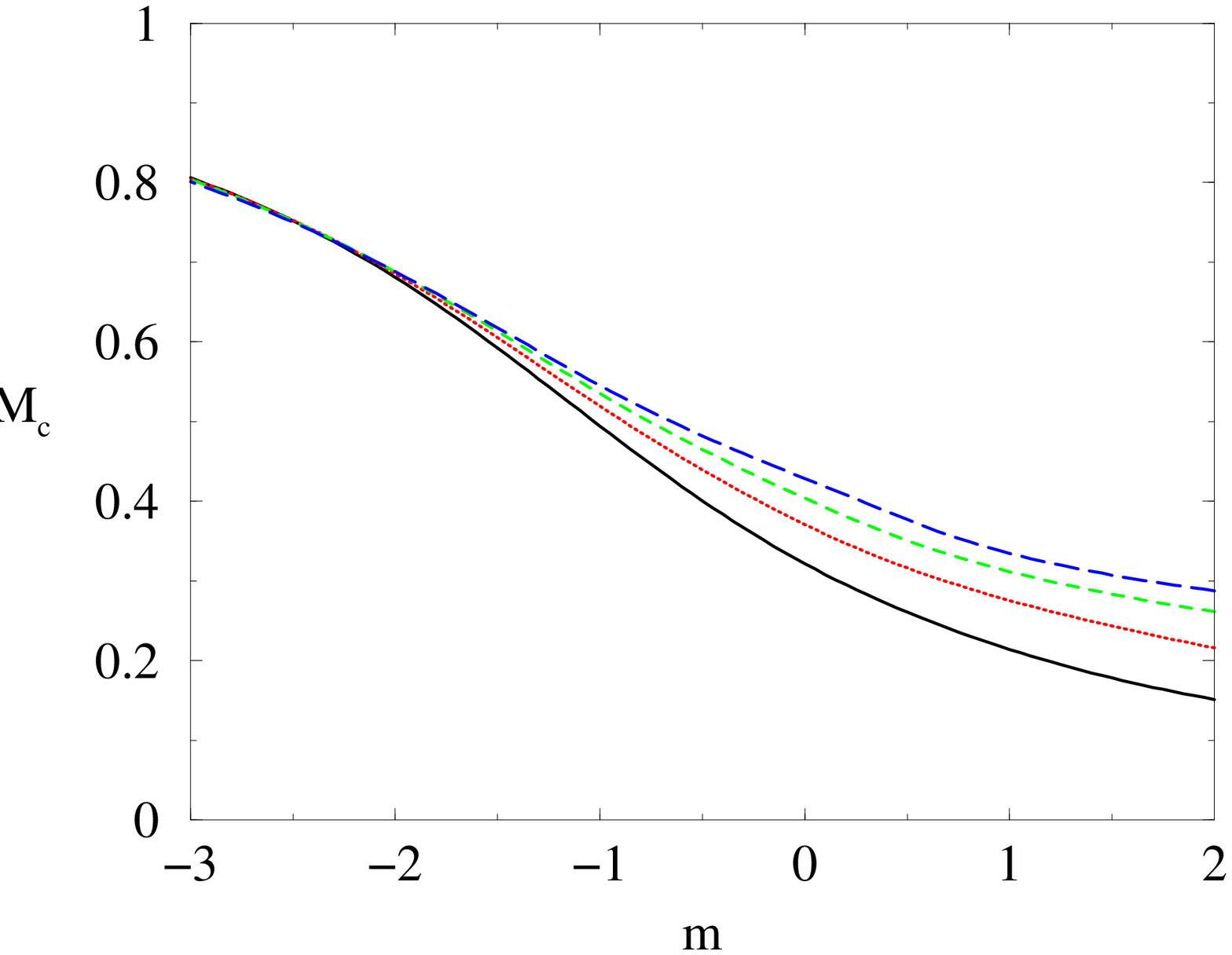}
\caption{Unconstraint case. Left panel: $\alpha_c$ versus biased parameter $m$ for different values of the threshold $\kappa=0.0,0.2,0.4$ and $0.6$ (from top to bottom).Right panel: The typical critical size $M_c$ of the subset as a function of $m$ for different values of the threshold $\kappa=0.0,0.2,0.4$ and $0.6$ (from bottom to top).}
\label{fig:unconstraintZE}
\end{center}
\end{figure}
We also studied the behavior of the system at $\alpha=\alpha_c$. The zero entropy condition (at zero temperature) gives the following equation for $\alpha_c$
\begin{eqnarray}
\alpha_c&=&-\left[\frac{1}{2}\big(M\widehat{M}+q\widehat{q}\big)+\int Dt\log\left(1+\exp^{-\frac{\widehat{M}+\widehat{q}}{2} -t\sqrt{\widehat{q}}}\right)\right]\\
&&\times\left[\int Dt\log H\left(\frac{\kappa+mM+t\sqrt{q}}{\sqrt{M-q}}\right)\right]^{-1}\label{ sp5}\,\,.
\end{eqnarray}
Adding this to the previous saddle-point equations (\ref{ fe1})-(\ref{ fe4}) and solving them numerically allows the study of the critical behavior for both the unconstraint and constraint case. Figure \ref{fig:unconstraintZE}  reports $\alpha_c$ for the unconstraint case and figure \ref{fig:constraintZE} for the constraint one. This figure can be used in principle in order to determine $N$ in order to be in the fault-free phase, since $P$ is given by the component and $m$ by the manufacturing process.
 
\subsection{Imperfect regime, $\alpha>\alpha_c$}
In this regime the entropy decreases with the temperature until $T\leq T_c$ where it becomes zero, with $T_c$ the freezing temperature. The fraction of errors is held constant in this interval, {\it i.e.}  $f(\beta_c)=f(\infty)$. The freezing temperature $T_c$ is therefore given by the zero entropy condition $s(\beta_c)=0$ and the fraction of errors $f$ then reads
\begin{equation}
 f= e^{-\beta_c}\int Dt\frac{1-H\left(\frac{\kappa+mM+t\sqrt{q}}{\sqrt{M-q}}\right)}{e^{-\beta_c}+(1-e^{-\beta_c}) H\left(\frac{\kappa+mM+t\sqrt{q}}{\sqrt{M-q}}\right)}
\label{ fraction_of_errors}
\end{equation}
Adding then this condition to the saddle-point equations (\ref{ fe1})-(\ref{ fe4}) fixes the temperature to $T_c$ and their numerical solution allows to evaluate the expression (\ref{ fraction_of_errors}) for the fraction of faulty functions. Notice that for $\alpha=\alpha_c$ the freezing temperature $T_c=0$ and therefore $f=0$.

\begin{figure}[h]
\begin{center}
\includegraphics[width=0.4\textwidth,height=0.4\textwidth]{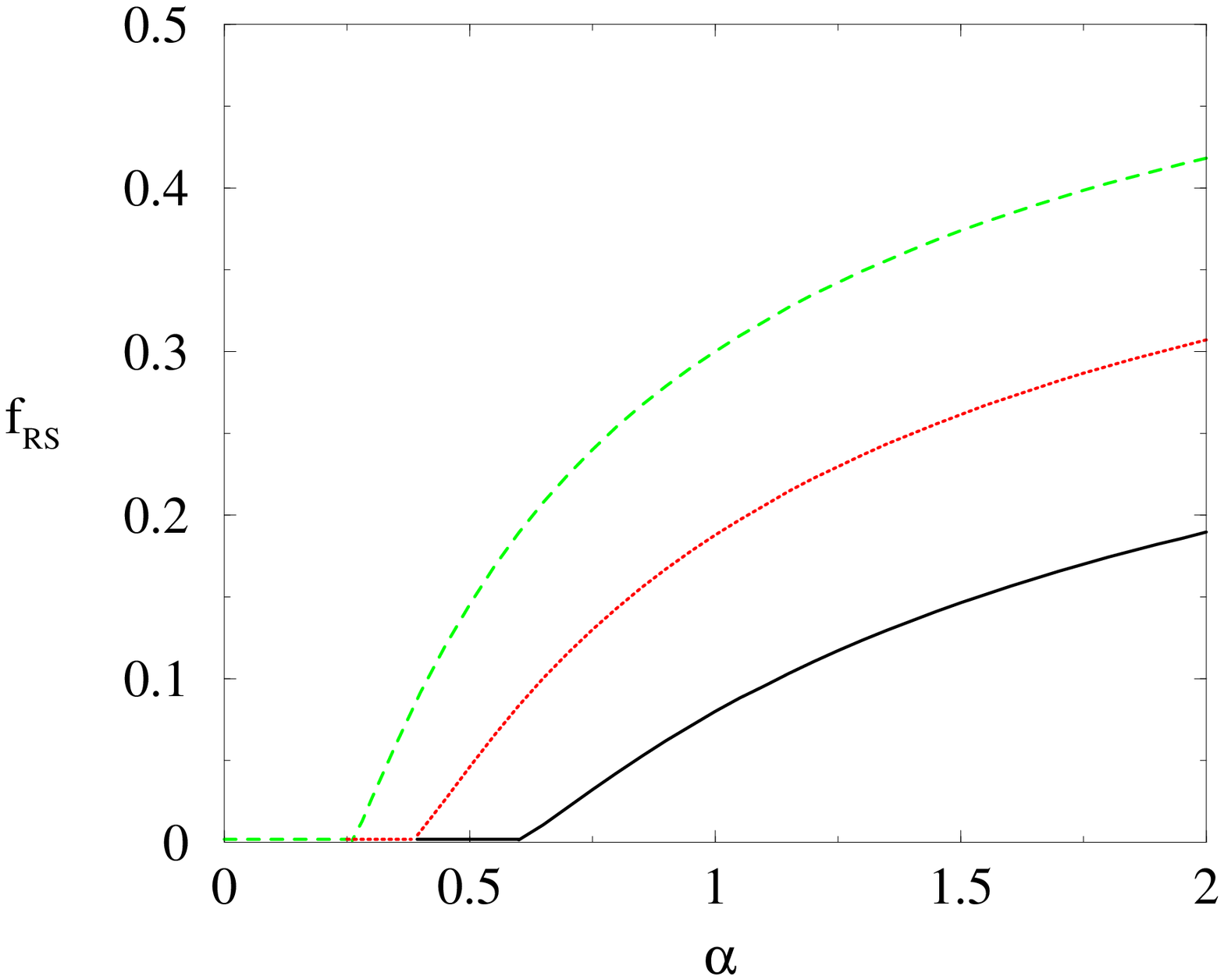}\quad 
\includegraphics[width=0.4\textwidth,height=0.4\textwidth]{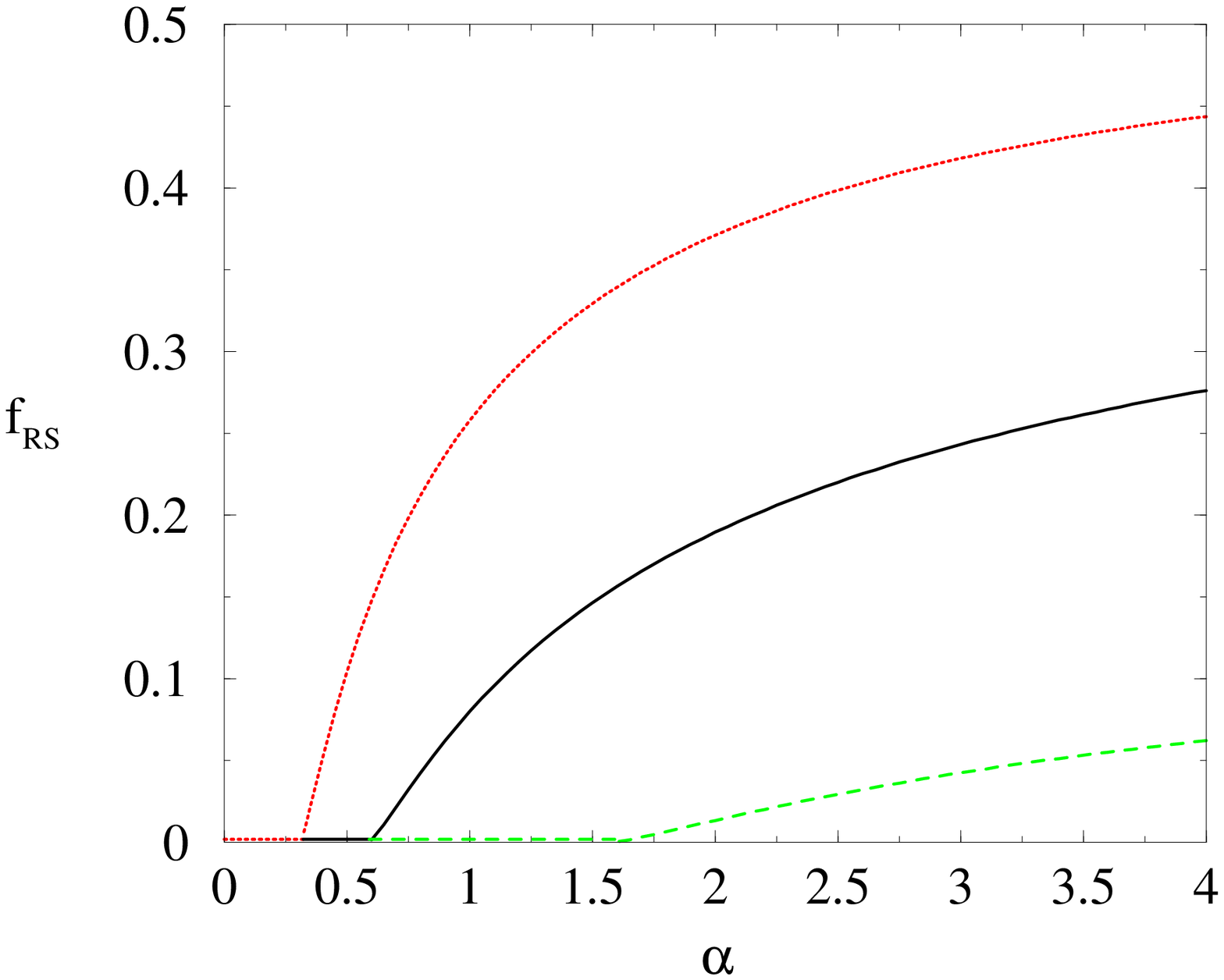}
\caption{Unconstraint case. Left panel: Fraction of faulty functions $f$ as a function of $\alpha$ for $m=0.0$ and $\kappa=0.0,0.2$ and $0.4$ (right to left). Right panel:  Fraction of faulty functions $f$ as a function of $\alpha$ for $\kappa=0.0$ and $m=-1.0,0.0$ and $1.0$ (right to left).}
\label{fig:error_unconstraint}
\end{center}
\end{figure}

\begin{figure}[h]
\begin{center}
\includegraphics[width=0.5\textwidth,height=0.4\textwidth]{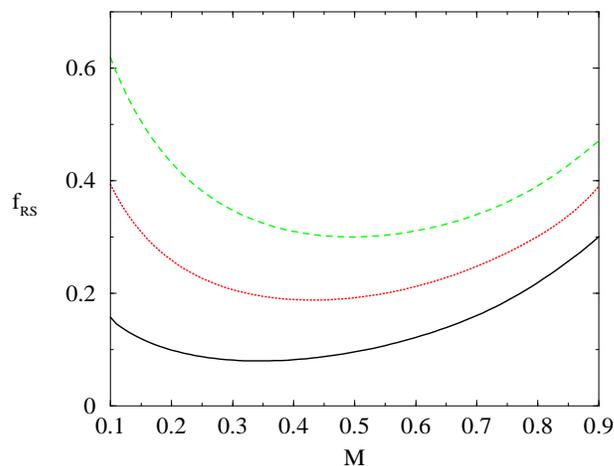} 
\caption{Constraint case. Fraction of faulty functions $f$ as a function of $M$ for $m=0.0$, $\alpha=1.0$ and  $\kappa=0.0,0.2$ and $0.4$ (bottom to top).}
\label{fig:error_constraint}
\end{center}
\end{figure}

Figure \ref{fig:error_unconstraint} shows the fractions of errors $f(\infty)$ versus $\alpha$ for different values of $\kappa$ and $m$, while in figure \ref{fig:error_constraint} we have fixed $\alpha$ and plotted the fraction of faulty functions against the subset size $M$.
\section{Numerical simulations}
\label{sec:simulations}
\begin{figure}[h]
\begin{center}
\includegraphics[width=0.5\textwidth,height=0.40\textwidth]{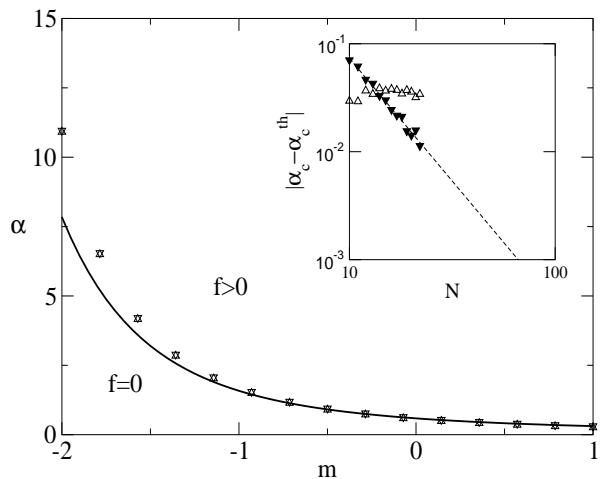}
\caption{Phase transition point $\alpha_c$ versus $m$. Theoretical results (continuous line), upper bound $\alpha_c^>$ (down triangles) and lower bound $\alpha_c^<$ (up triangles); $N=20$, average over $10000$ samples. Inset: finite-size analysis of $\alpha_c^>$ (solid down triangles) and $\alpha_c^<$ (up triangles) as a function of $N$ and for $m=0$; average over $10000$ samples.}
\label{fig:alphac_vs_m}
\end{center}
\end{figure}
\begin{figure}
\begin{center}
\includegraphics[width=0.5\textwidth,height=0.4\textwidth]{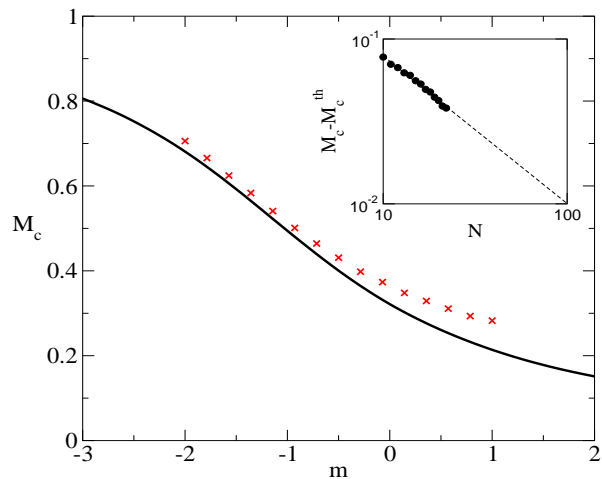}
\caption{Fraction $M_c$ of used components in the optimal combination at $\alpha_c$. $N=20$, average over 10000 samples. Inset: finite-size analysis of $M_c$ for $m=0$; average over $10000$ samples.}
\label{fig:Mc_vs_m}
\end{center}
\end{figure}
\begin{figure}
\begin{center}
\includegraphics[width=0.5\textwidth,height=0.4\textwidth]{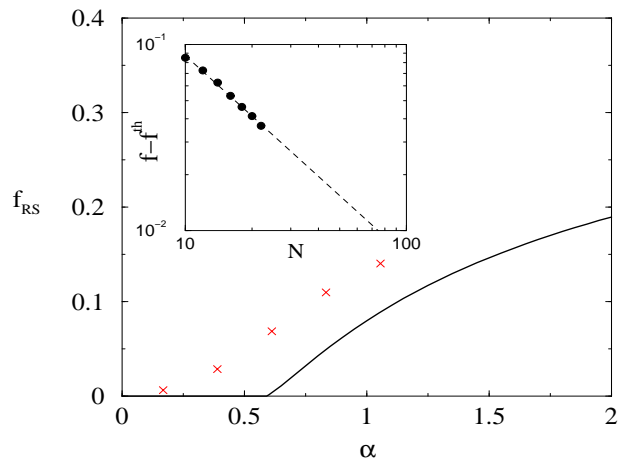}
\caption{Fraction  $f$ of working functions as a function of $\alpha$. $N=20$, average over 10000 samples. Inset: finite-size analysis of $f$ at $\alpha=0.5$; average over $10000$ samples.}
\label{fig:f_vs_alpha}
\end{center}
\end{figure}

We carried out extensive numerical simulations in order to check the above theoretical results. As in many similar neural network models, finite size effects are problematic in the binary DCP \cite{DerridaFinite,PRLcomment}. Figure \ref{fig:alphac_vs_m} plots $\alpha_c$ versus $m$ both for theoretical results, and numerical simulations. Fixing $N$ at 20, we enumerated all the combinations of components: starting with $P=1$ we added patterns, i.e. increased $P$, increasing thereby the set of disorder until $P=P^*$ for which no perfect combination can be found. The estimate of the critical point $\alpha_c(N)=P_c/N$ is $(P^*-1)/N\ge \alpha_c\ge P^*/N$, which gives lower and upper bounds for $\alpha_c$, noted $\alpha_{c}^<$ and $\alpha_{c}^>$ respectively. The agreement between theory and simulations is qualitatively good as long as $m$ does not take large negative values. We checked that the discrepancy between the numerical simulations and the theory is probably a finite size effect (see inset of figure \ref{fig:alphac_vs_m} for $m=0$): fitting $\alpha^>_c(N)$ first without imposing any asymptotic value $\alpha_c(\infty)$, i.e. with $\alpha^>_{c}(N)=aN^b+c$ yields $\alpha^>_c(N)=9.194N^{-2.093}+0.587$ with errors of $6.428$, $0.330$ and $0.006$ respectively. The error on $a$ is very large, while that on $c$ gives a surprisingly precise estimate of the theoretical value $\alpha_c=0.58976\dots$. Fitting our data with $\alpha_{c}^>(N)=aN^b+0.5898$ gives $\alpha^>_c(N)=11.602N^{-2.204}+0.5898$ with much smaller errors of $2.178$ and $0.075$ respectively. The lower bound $\alpha^<_c$ barely increases with $N$ and stays at around 0.55.. Many other quantities have the same kind of finite size scaling as $\alpha^>_c$, as, for instance, the fraction of used components $M_c$ at $\alpha_c$ (figure \ref{fig:Mc_vs_m}) and the fraction of faulty functions $f$ in the optimal combination (figure \ref{fig:f_vs_alpha}). A notable exception is that of the fraction of faulty functions in the constrained case. The integer nature of the problem causes notable variations depending on $M$ and $N$. Despite relatively large finite size effects, the numerical simulations confirm the validity of the theory.
\section{Flux recycling}
\label{sec:recycling}
The DCP studied above is static in nature, and does not address the whole complexity of component recycling, as in real life manufacturers produce a flux of faulty devices. How to recycle a flux is therefore a relevant problem. Let us start with some simple theoretical considerations. The central quantity of interest is the average quality of the components in the optimal combination (whose sense will be defined below), defined as the fraction of working functions of the components included in a combination $\bsigma$, i.e.
\begin{equation}
q(\bsigma)=\frac{1}{2}\left(1+\frac{\sum_{\mu=1}^P\sum_{i=1}^N\xi_i^\mu\sigma_i}{PNM(\bsigma)}\right)\,.
\end{equation}
Note that equation \req{condition} implies that the quality of a working combination is always greater than $1/2$ if $\kappa$ is fixed to 0. 

Let us assume that we have $N_0$ components initially and that $N_0$ and $\kappa$ are fixed so as to be in the fault-free region ($\alpha<\alpha_c$). In the following, we shall neglect fluctuations. The typical fraction of working functions is $w_0=q(\{1,\ldots,1\})=(1-m/\sqrt{N_0})/2$. If we now remove the optimal subset $\bsigma_0$ with, on average, $N_0M_0$ components of quality $q_0$, we have a new ensemble of $N_1=N_0(1-M_0)$ components with quality
\begin{equation}
w_1=(w_0-M_0q_0)/(1-M_0)\,.
\end{equation}
If new $N_0F_0$ fresh components taken from the flux of imperfect components are added, one has instead that the next iteration has $N_1=N_0(1-M_0+F_0)$ components with 
\begin{equation}
w_1=\frac{w_0-M_0q_0+F_0w_0}{1-M_0+F_0}
\end{equation}
and
\begin{equation}
\alpha_1=\frac{\alpha_0}{1-M_0+F_0}\,.
\end{equation}
Generalizing this equation to the $n$-th step yields
\begin{equation}
\label{ w_n+1}
w_{n+1}=\frac{w_n-M_nq_n+F_nw_0}{1-M_n+F_n}
\end{equation}
and 
\begin{equation}
\label{ alpha_n+1}
\alpha_{n+1}=\frac{\alpha_n}{1-M_n+F_n}\,.
\end{equation}
Flux recycling requires that the trajectory of $\alpha_n$ and $w_n$ stays in the fault-free region. The flux problem can be solved at constant $N_n=N$, that is, $F_n=M_n$, in which case Eq \req{ w_n+1} becomes
\begin{equation}
\label{ w_n+1_Ncst}
w_{n+1}=w_n+M_n(w_0-q_n)\,.
\end{equation}
\begin{figure}
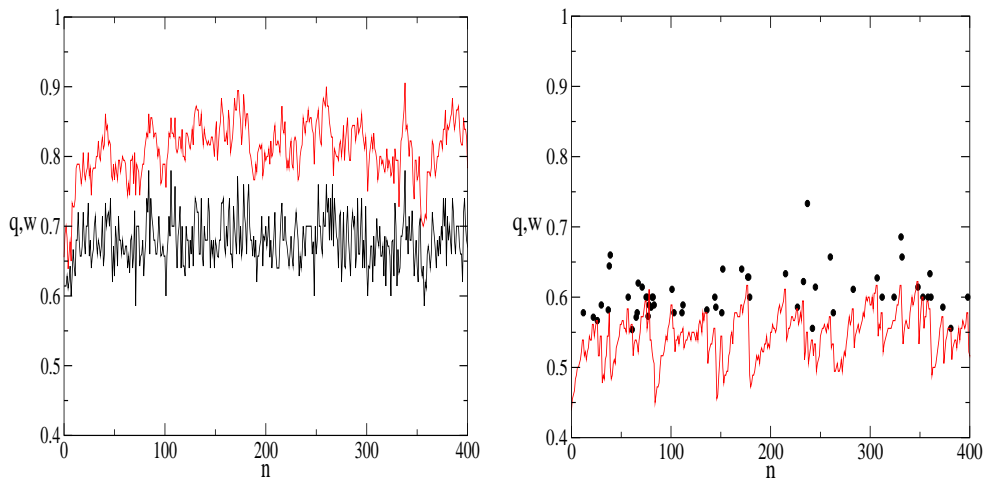

\begin{center}
\includegraphics[width=0.4\textwidth,height=0.4\textwidth]{figeps/qw_vs_n-m-1.5.eps}\quad
\includegraphics[width=0.4\textwidth,height=0.4\textwidth]{figeps/qw_vs_n-m0.eps}
\caption{Fraction of working functions $w$ in the set of imperfect components (red line) and quality $q$ of the optimal perfect combination (left graph: black line; right graph: circles) in dynamical flux recycling for $P=10$, $N=18$ and $m=-1.5$ (left graph) and $m=0$ (right graph).}
\label{fig:qw_vs_n-m-1.5}
\end{center}
\end{figure}
whereas $\alpha_{n+1}=\alpha_n$. We propose two main ingredients. First of all, in the faulty-free region, there is an exponentially large number of perfect combinations; which one is it best to select? In a static view, the one with the least number of components is the most economical. However, as suggested by equation \req{ w_n+1_Ncst}, $q_n$ should be minimized so as to make more probable that $w_{n+1}$ does not decrease as a function of time, which would inevitably lead the system out of the faulty-free region. Therefore, we define the optimal perfect combination as the one with the smallest $q$. A remarkable consequence of this choice is that this actually {\em increases} $w_n$ beyond $w_0$, as expected from the above discussion, and hence ensures that a perfect combination is found at each time step if $\alpha_0$ is sufficiently far from $\alpha_c$ (see figure \ref{fig:qw_vs_n-m-1.5}). Note as well that {\em no component is wasted}, hence the efficiency of this recycling scheme is 100\%. If $\alpha$ is smaller but close to $\alpha_c$, $\kappa$ can be adjusted dynamically (i.e. lowered if needed) to compensate for adverse fluctuations of $w_n$.

\begin{figure}
\begin{center}
\includegraphics[width=0.4\textwidth,height=0.4\textwidth]{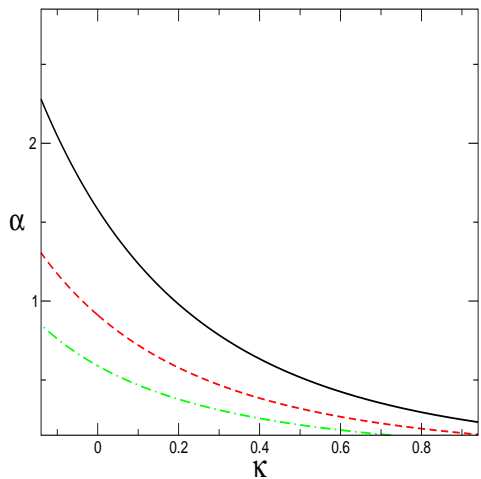}
\caption{Phase diagram for $m=0$ (continuous line), $-0.5$ (dashed line) and $-1$ (dot-dashed line)}
\label{fig:alpha_vs_K}
\end{center}
\end{figure}

When $\alpha$ is either close or above $\alpha_c$ and $N_n$ and $\kappa$ are kept constant, a new ingredient is needed. If no perfect combination is found, a simple but effective idea is to replace the worst component by a fresh one, until a perfect combination can be found. This keeps the recycling process going on forever, and makes it possible even for $\alpha>\alpha_c$. The price to pay is that some of the worst components will be wasted. Interestingly, the value of $w_n$ such that perfect combinations with average quality $q$ can be found is entirely determined by $\alpha$, i.e. independent from $m$. If we start at $\alpha>\alpha_c$, eliminating the worst components increases $w$ until $w\simeq q$ (see figure \ref{fig:qw_vs_n-m-1.5}).
Note however that $\alpha>\alpha_c$ can usually be avoided by lowering $\kappa$, unless the manufacturing process is really poor. Figure \ref{fig:alpha_vs_K} shows what $\kappa$ to choose for given $\alpha$ and $m$.

\section{Summary and conclusions}
\label{sec:conclusions}
In this paper we have solved the binary DCP at One-Step of RSB. The system is characterized by a phase transition ---similar to Random Energy model \cite{De81} or the Gardner capacity problem with Ising couplings \cite{Kr89}--- from a faulty-free regime with an exponential number of perfect subsets to an imperfect regime where no perfect combinations are available. We have contrasted our analytical findings with extensive numerical simulations based on exact enumeration. Even though they present strong finite size effects as in other models \cite{DerridaFinite,PRLcomment}, they show the validity of the theory qualitatively, but we can not rule out that in some regions further steps in the RSB are needed.

We have also addressed the dynamic problem of flux recycling and have proposed efficient methods that lead to no wastage at all.

\section*{Acknowledgment}
The authors thank the Abdus Salam ICTP for hospitally during the starting stages of this work. DC thanks Wadham College for support. IPC thanks the Fund for Scientific Research-Flanders, Belgium for financial support and R. Heylen for a careful reading of the manuscript.

\end{document}